\documentstyle[12pt,epsfig]{article}
\begin{document}
\begin{titlepage}
\vspace{.3in}
\begin{center}
{\Large{\bf The weight for random quark masses}}\\
\vskip 0.3in
{\bf John F. Donoghue}\\[.2in]
{\it Department of Physics and Astronomy, \\ 
University of Massachusetts,
Amherst, MA 01003 \\}

\end{center}
\vskip 0.4in

\begin{abstract}
In theories in which the parameters of the low energy theory are not
unique, perhaps having different values in different domains of the universe
as is possible in some inflationary models, the fermion masses would
be distributed with respect to some weight.
In such a situation the specifics of
the fermion masses do not have a unique explanation, yet the weight
provides the visible remnant of the structure of the underlying
theory. This paper introduces
this concept of a weight for the distribution of masses and provides
a quantitative estimate of it from the observed quarks and 
leptons. The weight favors light quark masses and appears roughly
scale invariant ( $\rho \sim 1/m$). Some relevant
issues, such as the running of the weight with scale and the possible
effects of anthropic constraints, are also discussed.

\end{abstract}
\end{titlepage}
\section{Basic ideas}

Many of the parameters of the Standard Model, such as the quark masses
and the weak mixing angles, appear without any obvious pattern.
Perhaps the explanation for the specific values of these parameters is
hidden in the physics at a deeper level. The goal of much of the work
in particle physics has been to find the underlying theory, the golden
Lagrangian, which by its structure explains the parameters of the
Standard Model. If this is successful, it will indeed be satisfying.

However, another possibility also exists - that these parameters are
in some sense random. This could potentially occur in various ways.
For example, in some theories of inflation, different regions of the
universe involve different parameters and perhaps even different low
energy theories[1]. The dynamical fields which determine the properties
of the low energy theory become fixed at different values in each
domain, and
subsequent inflation ensures that we live within only one of these
domains. It is also conceivable in theories such as superstrings
with different classically equivalent vacua, that different vacuum
states could be selected in different regions of the 
universe. The moduli fields, whose vacuum expectation values 
determine the mass parameters of the
low energy theory, are not fixed upon compactification[2]. Perhaps
these fields are sampled in a random fashion rather than being
determined uniquely by some mechanism. However, the general idea can
at this time be considered distinct from the particular underlying
theory.
If some quasi-random
mechanism is in fact at work, we should not expect to find a unique
explanation for the values of the parameters in the Standard Model.

Given our present incomplete knowledge, it is not any more scientific
to assume that there is a unique explanation for a single set of  
parameters which holds throughout the universe than it is to consider
the possibility that the parameters may vary in different domains. 
Indeed, the history of science has taught us that we do not occupy a
privileged position. In this spirit, it is fair to explore the
possibility that our domain of the Universe and our particular
parameters are not unique.

The quark masses do not appear strictly random either, as there are
more light quarks than very massive ones. Indeed, we would not 
necessarily
expect the masses to be equally distributed in such a multiple domain
theory. The structure and dynamics of the fields which determine the
low energy parameters may bias the distribution of possible
parameters in various ways. The parameters would then be distributed
randomly with respect to some weight. The potentially visible remnant
of the underlying theory would not be the details of the individual
mass values, but rather the weight by which they are distributed. If
we can obtain an indication of the weight, perhaps we can use this to
help determine the appropriate underlying theory.

The purpose of this paper is to provide an initial exploration of this
idea of a weight for fermion masses, and provide an estimate of the
weight from the observed masses. If we had available information from
an ensemble of different domains or even from very many masses within
a domain, the determination of the weight would be simple.
Because we are aware of only one
domain which contains six quark and three known lepton masses, we have
quite limited data for this exercise. However, the fermion mass
distributions are quite striking, and we can at least provide some
quantification of this fact.

The weight is not an invariant concept, identical at all scales.
Because the quark and lepton masses run under a change of scale, the
weight extracted at different scales will also run. I give a discussion and
quantification of this feature (Sec. 3), and present final masses at
the weak interaction scale (Sec 4). The smoothing procedure to obtain
information from a discrete spectrum is discussed in Sec 5.
Attempts to increase the number of
input parameters and to assess the uncertainty in this procedure 
are addressed in Sec. 6-8.
Further information is contained
in the Yukawa couplings that generate the weak mixing angles, although
the diagonalization of quark mass matrices lead to a loss of some of
this information(Sec 6). Lepton masses may also be relevant.
Since, quarks and leptons run
at different rates, there are also some complications in attempting to
combine these into a  single weight. This is explored using the
effects of QCD interactions (Sec 7). The uncertainties are summarized
in Sec 8, and then I try to provide a functional measure of the weight
in Sec 9. 

In a multiple domain theory, it is an obvious requirement that out of
all the possible domains we must find ourselves in a domain with
parameters
amenable to the development of life[3,4,5]. This restricts the space
of
allowed parameters somewhat, most especially the parameters whose
values might otherwise have to be fine-tuned. Weinberg[3] has
estimated
the tiny range of anthropically-allowed values of the cosmological
constant from the requirement that the universe expand at a rate which
allows galaxies to form. My collaborators and I [4] have estimated the
allowed range of value of the Higgs vacuum expectation value from the
anthropic need for complex atoms to exist and be formed in the
universe, and have suggested this as a possible explanation for the
unnatural closeness of the weak scale and the QCD scale. If the
fermion masses are also variable, there are anthropic constraints on
these also[4], favoring having a light first generation.
The subsequent generations appear to not have much impact on anthropic
constraints. However, this issue does potentially complicate attempts
to determine the weight, as I discuss later in the paper (Sec. 10).
At this stage, I also discuss further considerations which may modify
the procedure used to extract the weight appropriate for a given
underlying theory. Some of 
these issues will be addressed more
clearly if we are able to explore a specific underlying theory in an
attempt to predict the observed weight.

\section{Definition of the weight}
  
The weight provides a normalized distribution function for the masses
or Yukawa couplings. In an ensemble of domains similar to our own,
the fraction of masses found at a value $m$ within a range $dm$ is
defined to be 
\begin{equation}
 f(m) = \rho (m) ~dm
\end{equation} 
where $\rho (m)$ is the symbol for the weight. When we need to focus
on the scale dependence of the weight we will include the scale $\mu$
as a subscript, i.e. $\rho_{\mu} (m)$, indicating that it is the form
of $\rho$ appropriate for that value of $\mu$. By assumption, for a
small number of masses the values of the masses will appear randomly
distributed with respect to the weight $\rho (m)$.

The normalization of the weight is
\begin{equation}
 1 = \int \rho (m) ~dm \  \ .
\end{equation}  
The finiteness of the normalization imposes constraints on the low
mass and high mass limits of $\rho (m)$. At small values of $m$, the
weight cannot grow any faster than ${1 \over m}$ while for large
values it must fall faster than ${1 \over m}$. From this it is clear
that there is no pure power-law behavior that can hold for all values
of $m$.

\section{The running of the weight, quasi-fixed points, etc.}
   In the ultimate fundamental theory, the weight will be determined
by some physics at a large energy scale, for example the GUT or Planck
scales. In order for this to be compared with low-energy physics, it
will need to be transformed to a low-energy scale. Since the quark
masses run with the scale, the weight will similarly transform. This
section discusses the nature of the scale dependence of the weight.
    
The renormalization group equations provide a continuous 
one-to-one flow for the masses. In changing from a scale $\mu_1$ to a 
scale $\mu_2$ a mass $m_1$ will flow to a value $m_2$. 
This defines a functional relationship between 
the masses at the two scales, $m_1 = m_1(m_2)$ or $m_2 = m_2(m_1)$.
Similarly a small range of masses $\Delta m_1$ 
around the value $m_1$ will flow into a range $\Delta m_2$ 
around $m_2$. The magnitude of $\Delta m_2$ is generally not the same
as that of $\Delta m_1$ as a range of masses can either grow closer or
expand under the renormalization group transformation. The
transformation of the weight follows from a conservation of
probability. Since the transformation is continuous, the same fraction
of masses that fall in the range $\Delta m_1$ will, after the
rescaling, appear in the range $\Delta m_2$. From the definition of
this fraction as $f = \rho ~\Delta m$, we have the condition
\begin{equation}
   \rho_{\mu_2}(m_2) \Delta m_2  = \rho_{\mu_1}(m_1) \Delta m_1
\end{equation}
If we take the infinitesimal limit and define a ``Jacobian''
\begin{equation}
    J(m_2)  = {\partial m_1 \over \partial m_2}
\end{equation} 
as a function of $m_2$, we have the transformation equation
\begin{equation}
  \rho_{\mu_2}(m_2) = \rho_{\mu_1}( m_1(m_2)) J(m_2) \ .
\end{equation} 
The normalization is obviously preserved under this rescaling.

     As an example consider those masses where only the QCD gauge
couplings are important in the scaling of the masses. 
This implies a linear scaling with a simple anomalous dimension
\begin{equation}
m(\mu) = m(\mu_0) \left({\alpha_s(\mu) \over \alpha_s(\mu_0)}\right)^{d_m}
\end{equation}
with 
\begin{equation}
 d_m = {4 \over 11 - {2N_F \over 3}}
\end{equation}
In this situation we have the linear behavior 
\begin{eqnarray}
    m_2 &=& m_1 \left( {\alpha_s(\mu_2) \over \alpha_s(\mu_1)}\right)^{d_m} \\
  J(m_2) &=& \left( {\alpha_s(\mu_2) \over \alpha_s(\mu_1)}
   \right)^{-d_m} \\
   \rho_{\mu_2}(m_2) &=& \rho_{\mu_1}\left( m_2 ( {\alpha_s(\mu_2) \over
 \alpha_s(\mu_1)})^{-d_m}\right) \left( {\alpha_s(\mu_2) \over \alpha_s(\mu_1)
 }\right)^{-d_m} 
\end{eqnarray}
However, this simple form is only valid for small values of the
masses, as we now discuss. 

    When the mass, or equivalently the Yukawa coupling, is large the
running of the mass is influenced by the Yukawa coupling itself. When
evolving from high energy to low, the influence of the Yukawa coupling
tends to drive the mass to smaller values, while the gauge coupling
will tend to evolve the mass to larger values. If the Yukawa coupling
at the high energy scale is small enough, a finite evolution to lower
energy
will always be dominated by the gauge couplings, with the pattern
discussed in 
the previous paragraph. However, for larger initial
Yukawa couplings, the flow will approach a value where the effects of 
gauge and Yukawa couplings cancel. In particular all very large Yukawa
couplings will quickly approach this quasi-fixed point[6,7]. The effect of
this on the weight is that for reasonable distributions at high
energy, the low-energy weight will have an upper cutoff at the
quasi-fixed point.

Consider the evolution of a large 
Yukawa coupling $h$ under the influence of itself plus QCD interactions.
(We here neglect the weak and electromagnetic effects.)
Recall that within the Standard Model,
the masses are related to the Yukawa couplings $h_i$
via
\begin{equation}
m_i = {h_i \over \sqrt{2}} v 
\end{equation}
with $v = 246$ GeV being the Higgs vacuum expectation value. 
If $\mu_1$ is the initial scale and $t=ln(\mu_1^2/\mu^2)$, the
renormalization group equations for $N_f=6$ are[6,7]
\begin{eqnarray}
  {d g_3^2 \over dt} &=& {7 \over 16\pi^2} g_3^4 \\
  {d h^2 \over dt} &=& h^2 \left( {1 \over 2\pi^2} g_3^2 - {9 \over
32\pi^2} h^2 \right)
\end{eqnarray}
These equations have an exact solution[7]
\begin{eqnarray}
  h^2(t) &=& \left( {\alpha_s(t) \over \alpha_s(0)}\right)^{2d_m} 
{h^2(0) \over 1+ {9\over 8\pi} {h^2(0)\over \alpha_s(0)}\left(
({\alpha_s(t) \over \alpha_s(0)})^{1 \over 7} -1 \right)} \\
   &=&  {2 \over 9}
{g_3^2(t) \over 1+  
({\alpha_s(0) \over \alpha_s(t)})^{1 \over 7} \left( {2\over 9}
{g_3^2(0) \over h^2(0)} - 1 \right)}
\end{eqnarray}
The first form of this relation better illustrates the linear behavior
in the small $h$ limit, while the latter shows the quasi-fixed-point
at $h^2/g_3^2 = 2/9$. When scaling from the GUT or Planck scales
within the Standard Model, the quasi-fixed-point
occurs at $m^*=220$ GeV. This value would change if supersymmetry or
other interactions occurred in the region between the GUT and weak
scales.  
  
 The transformation of the weight follows directly from this form.
Let us define
\begin{eqnarray}
  h(t) &=& b {h(0) \over \left[ 1 + a h^2(0) \right]^{1 \over 2}} \\
  b(t) &=& \left( {\alpha_s(t) \over \alpha_s(0)}\right)^{4/7} \\
  a(t) &=& {9\over 2 g_3^2(0)} \left[\left({\alpha_s(t) \over 
              \alpha_s(0)}\right)^{1 \over 7} -1 \right] 
\end{eqnarray} 
We can invert this relation, obtaining
\begin{equation}
 h(0) = {h(t) \over \left[ b^2 - a h^2(t) \right]^{1 \over 2}}
\end{equation}
Recalling that $h(t)$ differs from the mass $m(t)$ only by a constant
(see Eq. (11)),
we find the Jacobian
\begin{equation}
  J(m) = {b^2 \over \left[ b^2 - 2a{m^2 \over v^2} \right]^{3 \over
2}} \ .
\end{equation}
For small masses, this is equivalent to Eq. 9.

    Let us explore the effects of rescaling and the quasi-fixed-point
via an example. At a high scale, such as the Planck mass, we consider 
\begin{equation}
   \rho_{\mu = M_P}(h) = {\lambda^{1-\delta} \over \Gamma (1-\delta)}
 {1 \over h^\delta} e^{-\lambda h} \ .
\end{equation} 
Finiteness constrains $\delta  < 1$. When scaled down to the weak
scale one obtains
\begin{equation}
   \rho_{\mu = M_W}(m) = {\lambda^{1-\delta} \over \Gamma (1-\delta)}
  \left({v \over \sqrt{2} m}\right)^{\delta} {b^2 \over \left[ b^2 -
2a{m^2 \over v^2} \right]^{3-\delta \over 2}} e^{-{\sqrt{2}\lambda m \over 
v\left[ b^2 - 2a{m^2 \over v^2} \right]^{1 \over 2}}} \ .
\end{equation}
This is plotted in Fig. 1 and 2 for $\delta = 1/2$ and $\delta = 0.9$
with $\lambda = 1$.
\begin{figure}  
\unitlength0.05in  
\begin{picture}(100,75)  
\put(0,0){\makebox(100,75)  
{\epsfig{figure=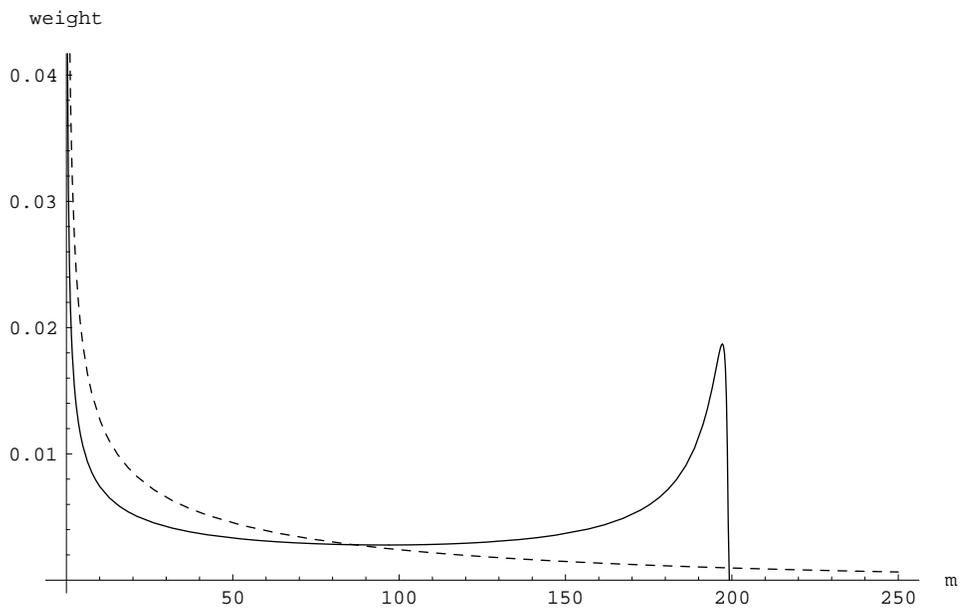,width=5in}}} 
\end{picture}  
\caption{The effect of renormalization group rescaling
on a possible weight
function with $\delta = 1/2$. The dashed curve corresponds to a weight
defined at the Planck scale and the solid curve is the same weight at
the scale $\mu = M_W$. } 
\end{figure}  
\begin{figure}
\unitlength0.05in
\begin{picture}(100,75)
\put(0,0){\makebox(100,75)
{\epsfig{figure=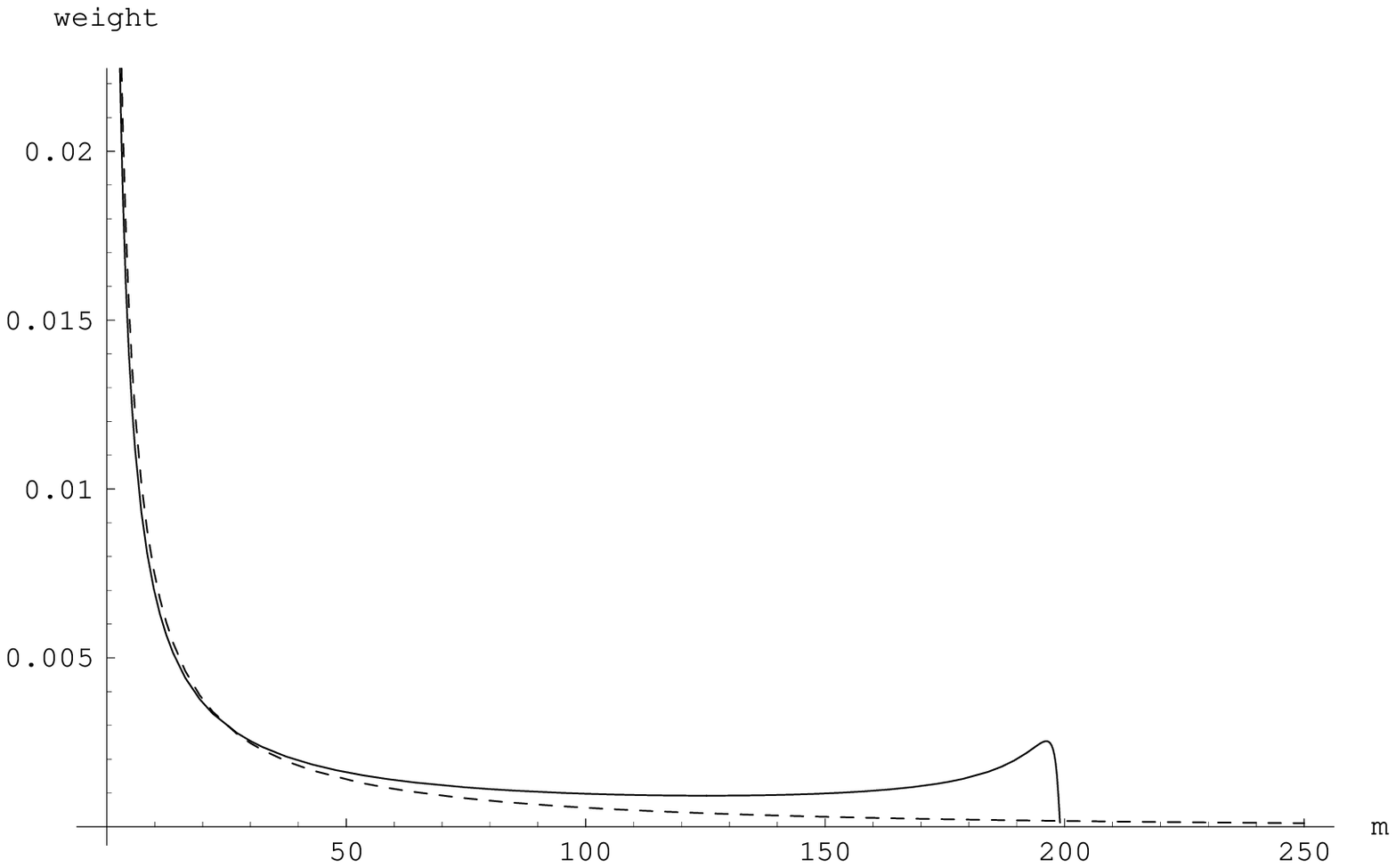,width=5in}}}
\end{picture}
\caption{The same as Fig. 1, but with $\delta = 0.9$.}
\end{figure}   
When scaled from the Planck mass to the weak scale, 
the parameters $a$ and $b$ have
values $a=7.9$ and $b=3.2$.
Note that while the high-energy distribution extends to large masses,
once rescaled to low energy, there is a cut-off at the
quasi-fixed-point. The integrated weight for large masses all appears
at values close to the fixed point, leading to a peaking of the
weight in this area. This indicates that even if the high-energy
weight is small for large $m$, it is nevertheless likely that one
or more of the masses will occur very close to the fixed point. It is
tempting to feel that this has occurred in the case of the top quark.

\section{Brief review of masses}

    The precise definition of quark masses involve many subtle issues,
which are not presently fully resolved. Because of confinement in
QCD, the quark masses can not be defined in the same way that we do for
leptons. The attempts to provide clear values of the masses involves
very interesting features. However, I will not dwell on many of 
these features
since this paper uses the masses only in a relatively crude way. I
will simply accept the uncertainties described in the review of Ref[8]. 

      The ratios of light quark masses are better determined than the
absolute magnitude. The ratios follow from the masses of pseudoscalar
mesons when analyzed using chiral methods[9]. To specify the magnitude, 
one must specify a specific renormalization condition. This is
presently only possible in model-dependent methods, leading to at
least a factor
of two uncertainty in the scale of the masses. We use
\begin{eqnarray}
 m_u &=& 4 ~{}^{+4}_{-2} \ MeV \\
 m_d &=& 8 ~{}^{+7}_{-4} \ MeV \\
 m_s &=& 150 ~{}^{+150}_{-50}   \ MeV
\end{eqnarray}
These
values are intended to be estimates in a mass-independent
renormalization scheme such as $\bar{MS}$ at a scale of order 1 GeV.

    Heavy quark masses can be defined to some level of precision in
the context of Heavy Quark Effective Theory (HQET)[10]. Here one can
define either a pole mass or a running mass evaluated at the scale of
the mass itself. For the running masses 
we use
\begin{eqnarray} 
m_c(m_c) &=& 1.4 \pm 0.2  \ GeV \\
m_b(m_b) &=& 4.3 \pm 0.2  \ GeV \\
m_t(m_t) &=& 166 \pm 5  \ GeV. 
\end{eqnarray}

In describing the weight, we need to transform the masses to a common
scale. I will primarily use $M_W$ as this scale. Running the masses to
this value yields
\begin{eqnarray}
 m_u(M_W) &=& 2.2 ~{}^{+2.2}_{-1.1}  \ MeV \\
 m_d(M_W) &=& 4.4 ~{}^{+4}_{-2}  \ MeV \\
 m_s(M_W) &=& 80 ~{}^{+80}_{-30} \  MeV \\
 m_c(M_W) &=& 0.81 \pm 0.12  \ GeV \\
 m_b(M_W) &=& 3.1 \pm 0.2  \ GeV \\
 m_t(M_W) &=& 170 \pm 5  \ GeV 
\end{eqnarray}
In this calculation I used $\alpha_s(M_W) = 0.115$ and included the
changes in $N_F$ at charm and beauty thresholds.
The
set of masses are equivalent to a set of dimensionless Yukawa
couplings. Recall that the top quark mass is equivalent to $h_t = 1$.

\section{Smoothing the quark distribution}
     With only six quarks our insight into the weight is necessarily
limited. In particular, there is nothing that we can do that can give
much information at large values of the mass. The single very heavy
quark, the top quark, tells us that the weight cannot vanish at large
mass, nor be exponentially suppressed. 
However it is not possible to use the top quark to say anything
detailed about the shape of the weight. The situation is marginally
better at low mass. Viewed globally there is a striking clustering of
the masses at low values. This requires the weight to be peaked at
low mass. If we assume a smooth functional form for the
weight, we will be able to compare it with some of the features of
the observed masses. So we should be aware that our limited
information on the weight exists almost entirely in the region
where most masses are, i.e. around and below a GeV.

In a very crude form, we can consider binning the masses to give a
rough estimate of the weight. This is illustrated in Fig. 3.
\begin{figure}  
\unitlength0.05in  
\begin{picture}(100,75)  
\put(0,0){\makebox(100,75)  
{\epsfig{figure=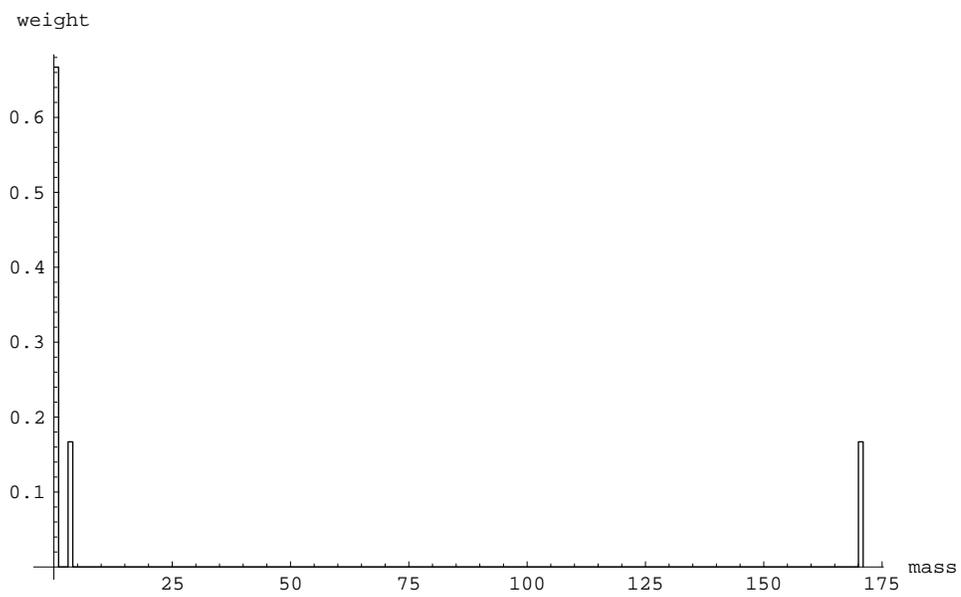,width=5in}}} 
\end{picture}  
\caption{A rough characterization of the weight, obtained by
binning the quark masses in 1 GeV bins. } 
\end{figure}  

 In order to compare with the physical masses in a more quantitative
fashion we need some
smoothing scheme which is able to encode the information contained in
the discrete values of the masses, yet which can be compared to possible
smooth trial functions describing the weight. I will use two such
schemes. One involves an integral with some
similarity to the Hilbert transform
\begin{equation}
H(z) = \int^\infty_0 dm ~{z \rho (m) \over m + z}
\end{equation}
The second involves the Laplace transform
\begin{equation}
L(z) = \int^\infty_0 dm ~\rho (m) ~e^{-m/z}
\end{equation}
These transforms are constrained by the normalization condition to
both be equal to unity at $z=\infty$ and are mainly sensitive to
masses smaller in magnitude than the parameter $z$. I will refer to
these functions as ``transformed weights''. The transformed weights
H(z) and L(z) turn out contain quite similar information, and I will
display L(z) only occasionally in what follows.

For the
experimental side we use
\begin{equation}
\rho_{exp} (m) = {1 \over 6}\Sigma^6_{i=1} \delta(m - m_i)
\end{equation}
The resultant transforms are displayed in Fig 4,5.
\begin{figure}
\unitlength0.05in
\begin{picture}(100,75)
\put(0,0){\makebox(100,75)
{\epsfig{figure=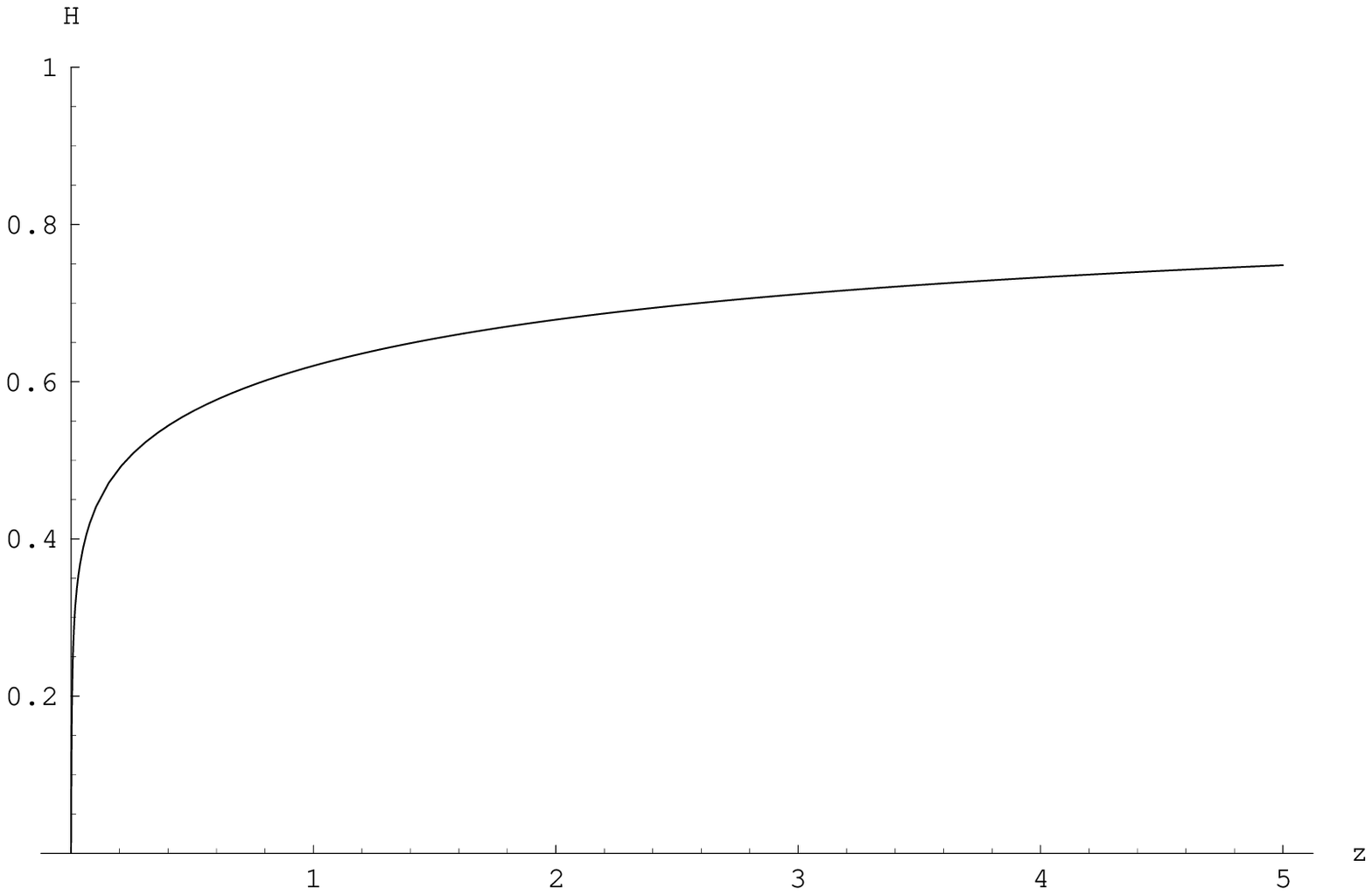,width=5in}}}
\end{picture}
\caption{The ``experimental'' transformed weight H formed from
the quark masses defined at the scale $\mu = M_W$. }
\end{figure}
\begin{figure}  
\unitlength0.05in  
\begin{picture}(100,75)  
\put(0,0){\makebox(100,75)  
{\epsfig{figure=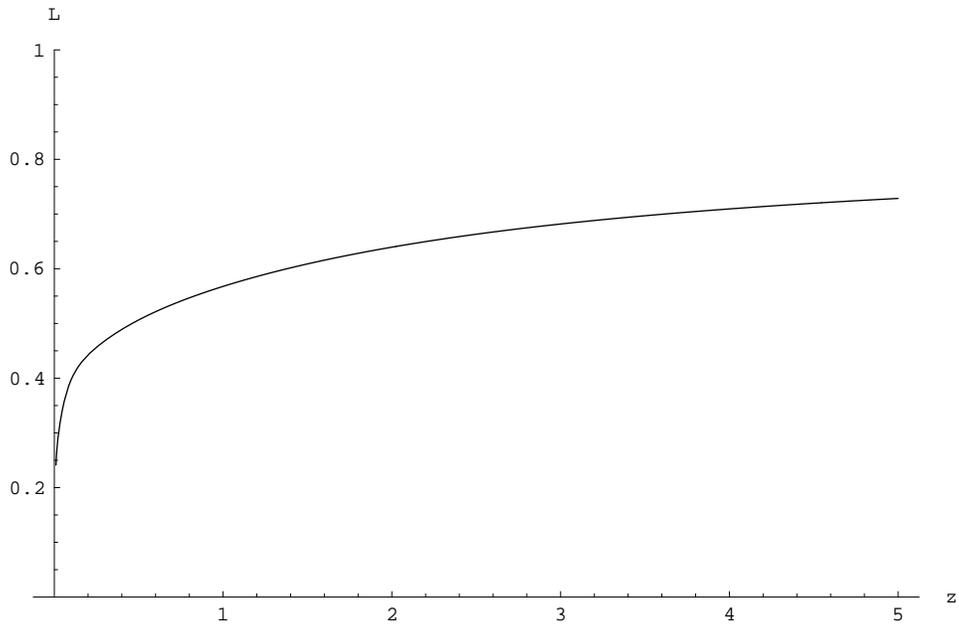,width=5in}}} 
\end{picture}  
\caption{The ``experimental'' transformed weight L. } 
\end{figure}  
  
Here and in subsequent sections, I wish to provide some estimates of
the uncertainty in the transformed weights. A minimal such estimate
is obtained by considering what would occur if we had knowledge of
only five quarks instead of six. Removing the information on the mass
of a strange or charmed quark does relatively little to change the
transformed weight. The extremes occur if we discard either the up or
the top quark masses. These shifts in the form of H(z) are illustrated
in Fig. 6. The effects of the error bars from the experimental
determination of the masses is much smaller than the effect of
removing one mass from the distribution. 
\begin{figure}  
\unitlength0.05in  
\begin{picture}(100,75)  
\put(0,0){\makebox(100,75)  
{\epsfig{figure=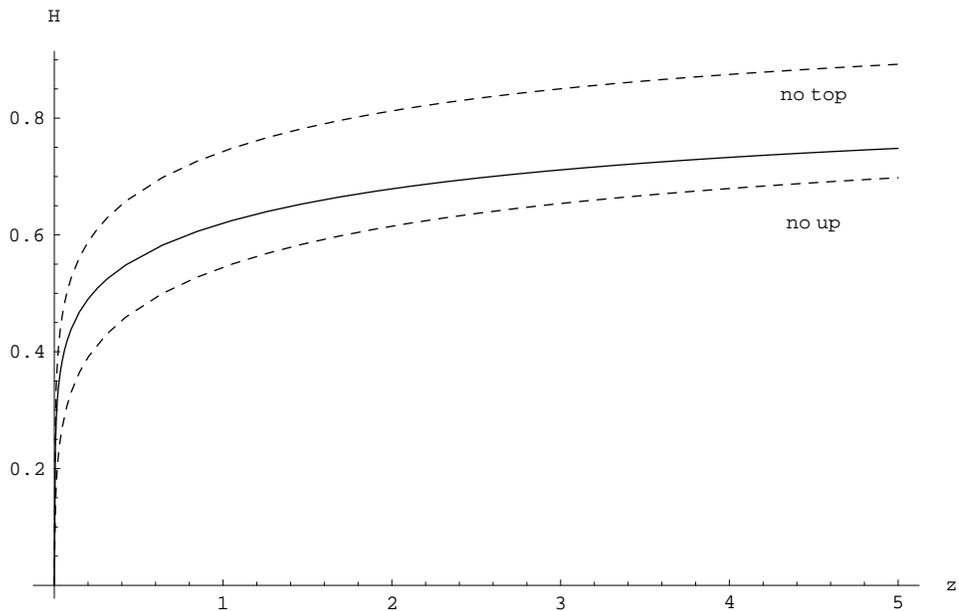,width=5in}}} 
\end{picture}  
\caption{The transformed weights obtained by discarding either
the up quark or the top quark, compared to the full result (solid
curve). } 
\end{figure}  

\section{The CKM weak mixing matrix}

The known quark masses emerge from the diagonalization of the matrix
of Yukawa couplings. This diagonalization
also produces the elements of the weak mixing matrix. Therefore the
CKM elements also contain information on the distribution of the
Yukawa couplings. In this section, I consider this information,
primarily as an indication of the uncertainty in the weight.

  There is a serious loss of information in the diagonalization
process which hinders our use of the CKM elements. The original mass
matrices of the charge $2/3$ and $-1/3$ quarks are diagonalized via 

\begin{eqnarray}
V_L^{(u) \dagger} M_0^{(u)} V_R^{(u)} & = & m^{(u)}    \\
V_L^{(d) \dagger} M_0^{(d)} V_R^{(d)} & = & m^{(d)}   
\end{eqnarray}
with the resultant CKM matrix formed from the product of the two
left-handed rotations
\begin{equation}
V_{CKM}  = V_L^{(u)\dagger}V_L^{(d)}  .
\end{equation}
We lose any information contained in the right-handed rotations and
any common features of the up and down type left-handed rotations. Thus
we cannot reconstruct the original Yukawa matrices except in an
arbitrary choice of basis. (Indeed, the choice of basis can only be
specified with reference to some physics beyond the Standard Model.)

In order to provide two simple estimates of the uncertainties in the
distribution of Yukawa couplings, I will rotate the CKM elements
either into the matrix of up-type quarks or into the matrix of
down-type quarks. Specifically I consider the elements of 
\begin{equation}
M_{u,test} = V_{CKM}^\dagger m^{(u)} =\pmatrix{0.002 & 0.18 & 1.19 \cr
0.0005 & 0.78 & 6.8  \cr 7 \cdot 10^{-6}  & 0.032  & 170 }
\end{equation}
or of 
\begin{equation}
M_{d,test} = V_{CKM} m^{(d)}  =\pmatrix{0.004 & 0.018 & 0.01 \cr
0.001  & 0.077 & 0.12 \cr 3 \cdot 10^{-5} & 0.003 & 3.1 }
\end{equation}
In the numbers quoted, I do not include CP violation and use the PDG
central values[8]. For the purpose of this exercise, I will consider
all elements to be independent. In each case, I also include the
diagonal masses of the other type quark to be included in the weight,
resulting in 12 input parameters per case. This leads to the
transformed weights shown in Fig 7.
\begin{figure}  
\unitlength0.05in  
\begin{picture}(100,75)  
\put(0,0){\makebox(100,75)  
{\epsfig{figure=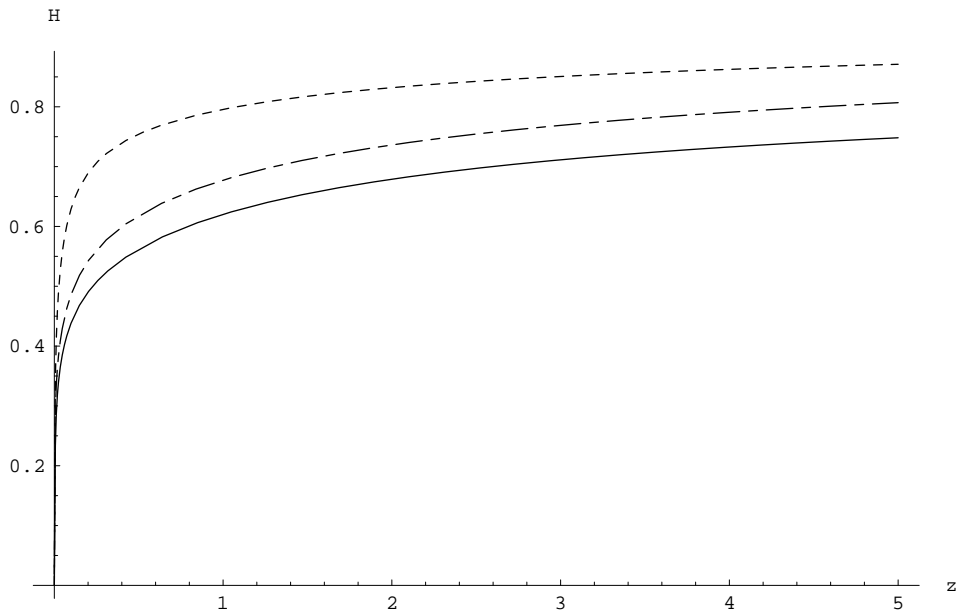,width=5in}}} 
\end{picture}  
\caption{The transformed weight (solid curve) compared to the
corresponding result formed by transferring the information in the CKM
elements into the up-quark mass matrix(dot-dashed curve) and
down-quark mass
matrix(dashed curve). } 
\end{figure}

We see that including this increased amount of data does not
significantly change the weight.

\section{Leptons}

It is tempting to also combine information of the lepton masses with
that of quark masses. Certainly the overall impression is the same,
with a preponderance of light fermions. However, we should be cautious
in this procedure as there are certainly some different considerations
for leptons compared to quarks. In the spirit of exploring the
uncertainties in the weight, in this section we combine quark and
lepton information. We will see that within the significant
uncertainties the quark and lepton distributions are
consistent. 

Initially let us blindly add the lepton to the quark information (i.e.
yielding 9 elements of data) This is shown in Fig. 8. (I neglect the
electroweak running of masses as this uncertainty is far below the
real uncertainty in this whole procedure.) Leptons produce only a
minor change in the transformed weight.
\begin{figure}  
\unitlength0.05in  
\begin{picture}(100,75)  
\put(0,0){\makebox(100,75)  
{\epsfig{figure=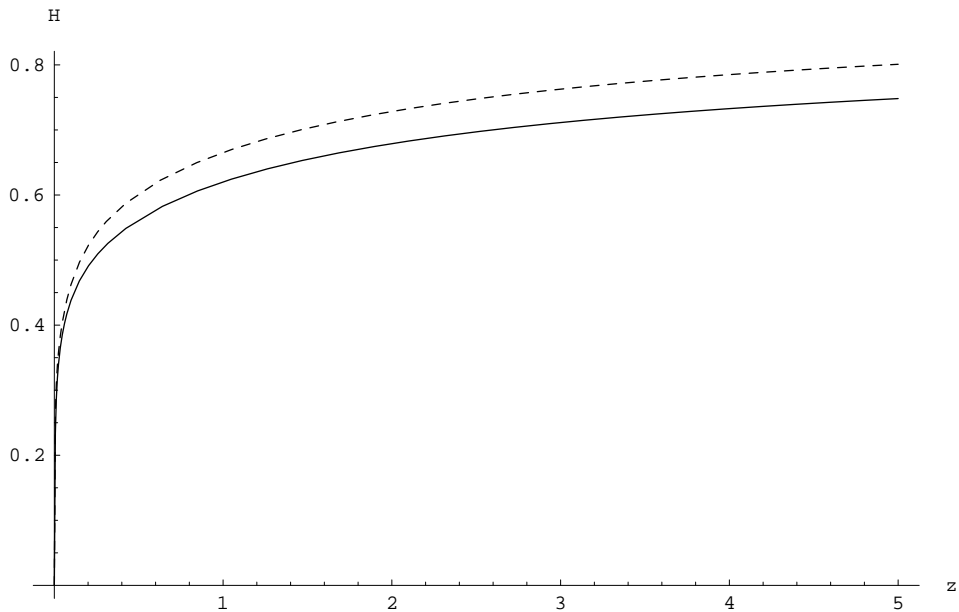,width=5in}}} 
\end{picture}  
\caption{The transformed weight formed from the quark masses
and the lepton masses (dashed curve), compared to that from quarks
alone (solid curve). } 
\end{figure}

However, if the masses arise at some higher energy scale, this is
unlikely to be the correct procedure. The quark and lepton masses
should be compared at the high scale rather than the electroweak
scale. This of course cannot be done without knowledge of the
underlying theory. To estimate this effect, I have done the following.
First, imagine that all masses are scaled to a typical grand
unification scale, $M \sim 10^{16}$GeV. Then all masses are scaled
back to the W scale as if they were quarks. This gives a plausible
common distribution, which can be compared to the result of the
previous section. The net effect is to increase the lepton
input values by a
factor of two. This change does not drastically modify the resulting
distribution. See Fig. 9. The lepton information is consistent with
that of the quarks.
\begin{figure}  
\unitlength0.05in  
\begin{picture}(100,75)  
\put(0,0){\makebox(100,75)  
{\epsfig{figure=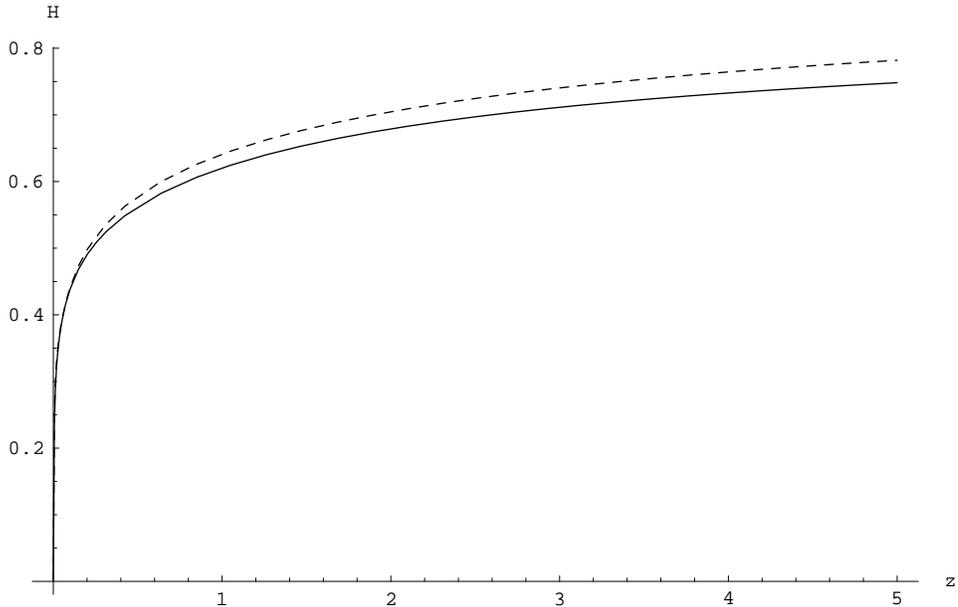,width=5in}}} 
\end{picture}  
\caption{The transformed weight formed from quark masses and
lepton masses which have been rescaled as described in the text
(dashed curve) compared to that from quarks alone (solid curve). } 
\end{figure}  

Neutrinos may or may not have a mass in the Standard Model. However,
if they have a non-zero mass, present indications are that the values
are so small that they are unlikely to be standard Dirac masses. The
``see-saw'' mechanism naturally explains such small masses and would
be the favored explanation should present indications be confirmed.
This mechanism would indicate that we should not combine neutrino mass
information together with the other masses in discussing the weight.

Finally, we combine the CKM, (rescaled) lepton and quark information in
order to obtain the estimate in Fig. 10.
\begin{figure}  
\unitlength0.05in  
\begin{picture}(100,75)  
\put(0,0){\makebox(100,75)  
{\epsfig{figure=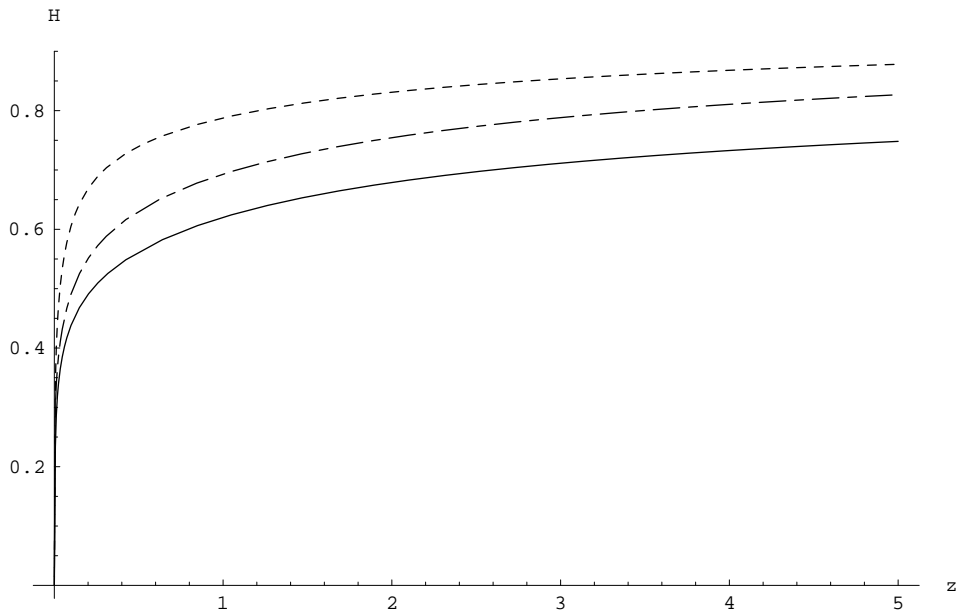,width=5in}}} 
\end{picture}  
\caption{A summary of all inputs (quarks, leptons and CKM
elements), with the CKM elements in the up-quark mass matrix
(dot-dashed curve) and the down-quark mass matrix ( dashed curve). } 
\end{figure}

\section{Summary of estimated uncertainties}

The considerations above have described various inputs into the
``experimental'' determination of the transformed weight. The number
of inputs has ranged from 5 to 15. The greatest downward 
variation comes from
the removal of the up quark from the
distribution. The upper boundary comes from the rotation of the CKM
elements into the down-quark mass matrix. All 
other inputs are consistent with
the underlying distribution within this uncertainty. 

The resulting uncertainty is shown in Fig. 11. Despite the generous
nature of the uncertainty in the transformed weight, it will turn out
that the weight itself is reasonably constrained.
\begin{figure}  
\unitlength0.05in  
\begin{picture}(100,75)  
\put(0,0){\makebox(100,75)  
{\epsfig{figure=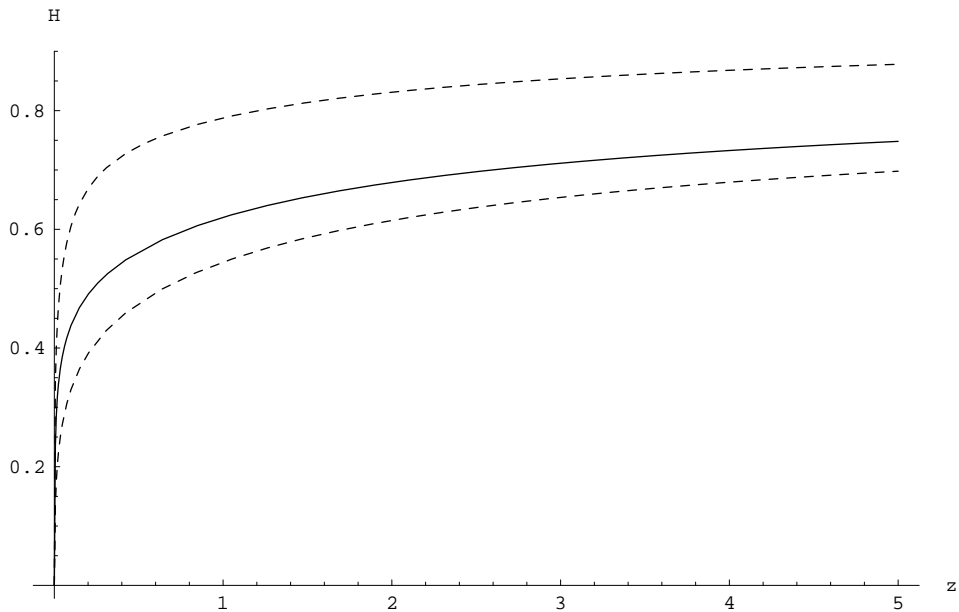,width=5in}}} 
\end{picture}  
\caption{The transformed weight (solid curve) and limits of
the range
of uncertainty (dashed curves) as described in the text. } 
\end{figure}

\section{Phenomenology of the weight}

The function describing the weight needs to be peaked at low energy,
yet extend out to high mass in order to accommodate the existence of
the top quark. The latter requirement eliminates purely exponential
forms which would have a scale of order 1 GeV, and hence no
significant probability at multi-GeV masses. Power law forms will be
seen to provide acceptable distributions, and we will explore some
variants of these.

Consider a trial weight of a pure power behavior combined with a
cutoff at the quasi-fixed point
\begin{equation}
\rho_1(m) = {N \over m^\delta} \Theta( m^* - m)
\end{equation}
with a normalization constraint $N=(1-\delta )/ m^{*(1-\delta)}$. 
I will use the cutoff at $m^* = 220$ GeV. Here we are constrained by
$\delta < 1$ for the distribution to be integrable at low mass.
This weight is shown in Fig. 12 for the values $\delta =
0,0.1,0.2,.....,0.9$. 
\begin{figure}  
\unitlength0.05in  
\begin{picture}(100,75)  
\put(0,0){\makebox(100,75)  
{\epsfig{figure=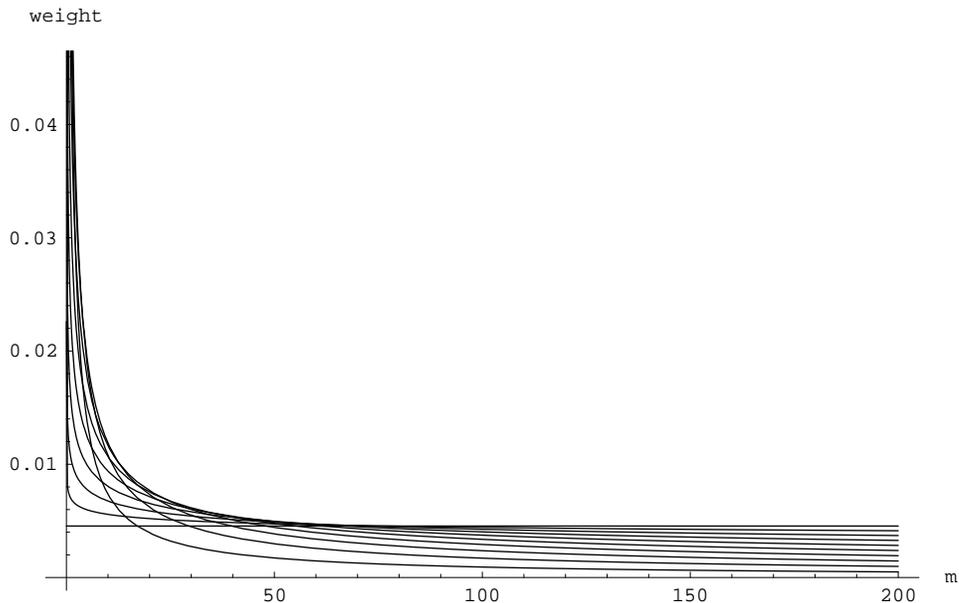,width=5in}}} 
\end{picture}  
\caption{Sample power-law weights, with powers ranging from
$\delta = 0$ to $\delta = 0.9$ } 
\end{figure}

The simplest consideration is the use of the median value of the
distribution, $\hat{m}$, defined by
\begin{equation}
{1 \over 2} = \int_0^{\hat{m}} \rho (m)~ dm
\end{equation}
On the average, half of the quark masses should appear below
$\hat{m}$. For this simple example 
\begin{equation}
\hat{m} = {m^* \over 2^{1 \over (1-\delta)}}
\end{equation}
Even a 
crude estimate 
\begin{equation}
 0.05 ~MeV \le \hat{m} \le 5 ~GeV
\end{equation}
leads to a rather stringent constraint on the power $\delta$,
\begin{equation}
0.82 \le \delta \le 0.955
\end{equation}
with higher values of $\delta$ corresponding to lower values of
$\hat{m}$. 

Similar constraints follow from even a rough look at the transformed
weight. In Fig 13 are shown the transformed weights for these
distributions for the same range of $\delta$.
Again, only higher values of $\delta$ are allowed. The physics behind
this is clear - if $\delta$ is small we would expect to populate
larger values of the masses, and hence would see the variation in
H(z) occur at larger values of z. A search for the optimal value of
$\delta$ leads to the trial with $\delta= 0.91$ shown in Fig 14. A
similar comparison of the same trial weight with the transform L(z)
yields an equally reasonable form, shown in Fig. 15. Of course, given
the uncertainties described above the specific value of this exponent
should not be taken too seriously. Within the scope of this class of
trial functions, I estimate that $\delta = 0.82$ to $\delta = 0.95$
spans the uncertainties described in the previous sections.
\begin{figure}  
\unitlength0.05in  
\begin{picture}(100,75)  
\put(0,0){\makebox(100,75)  
{\epsfig{figure=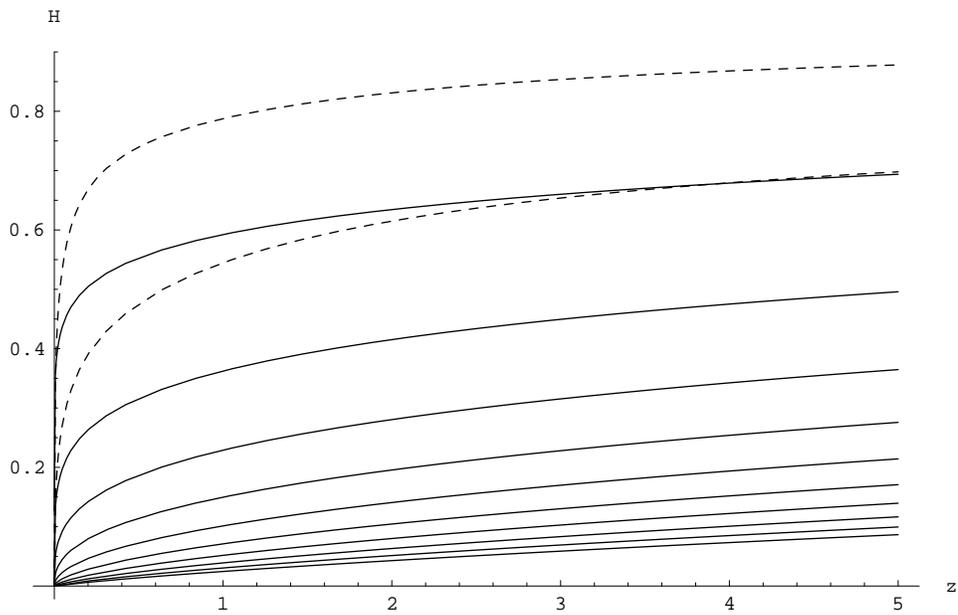,width=5in}}} 
\end{picture}  
\caption{The transformed weights corresponding to power-law
weights, with powers ranging from $\delta = 0$ (bottom) to $\delta =
0.9$ (top). Also shown is the range of uncertainty in the transformed
weight from Fig. 11. } 
\end{figure}  
\begin{figure}  
\unitlength0.05in  
\begin{picture}(100,75)  
\put(0,0){\makebox(100,75)  
{\epsfig{figure=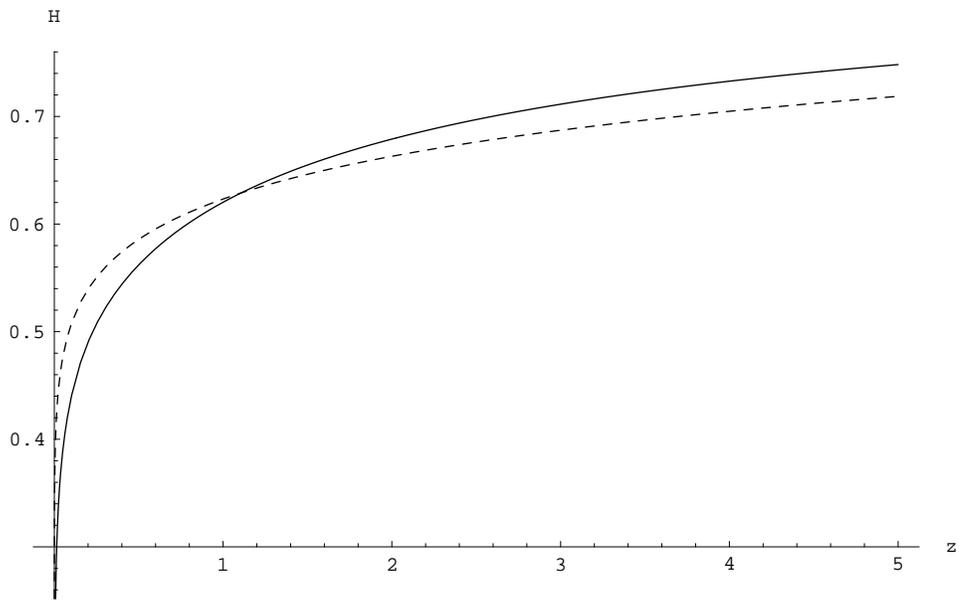,width=5in}}} 
\end{picture}  
\caption{A comparison of the experimental transformed weight
H with a power-law fit (dashed curve) with $\delta = 0.91$. } 
\end{figure}  

\begin{figure}  
\unitlength0.05in  
\begin{picture}(100,75)  
\put(0,0){\makebox(100,75)  
{\epsfig{figure=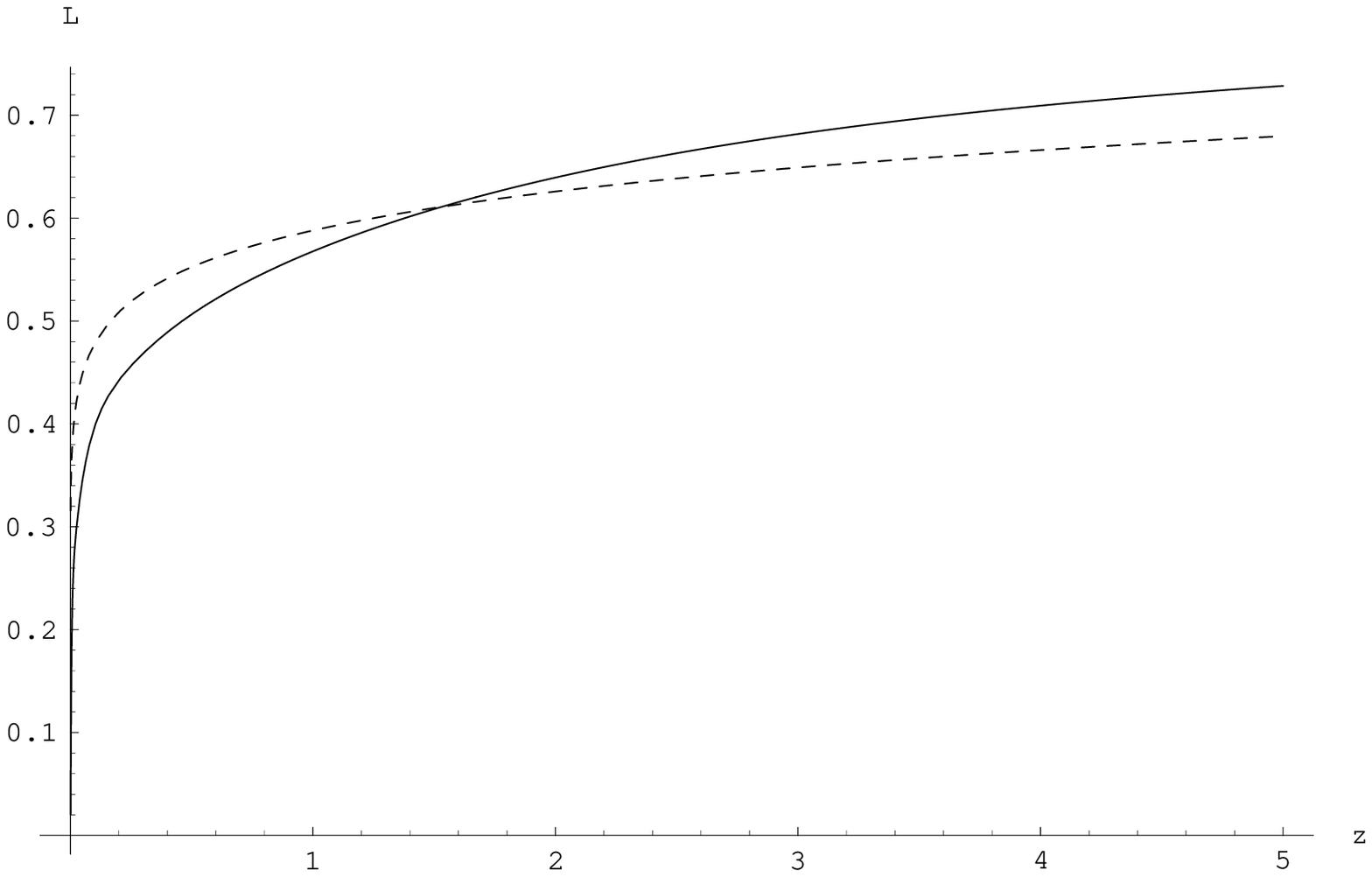,width=5in}}} 
\end{picture}  
\caption{The same as Fig. 14 but for the transformed weight L. } 
\end{figure}

A second class of trial functions consists of the form given in Eq. 22
from the renormalization group scaling down from high energies. This
differs primarily in the region of the high mass cut-off. 
Here a similar power provides a similarly good fit, as shown in Fig 16
for $\delta=0.92$. This confirms the expectation that the high energy
form of $\rho (m)$ is not significantly constrained by the data.
\begin{figure}  
\unitlength0.05in  
\begin{picture}(100,75)  
\put(0,0){\makebox(100,75)  
{\epsfig{figure=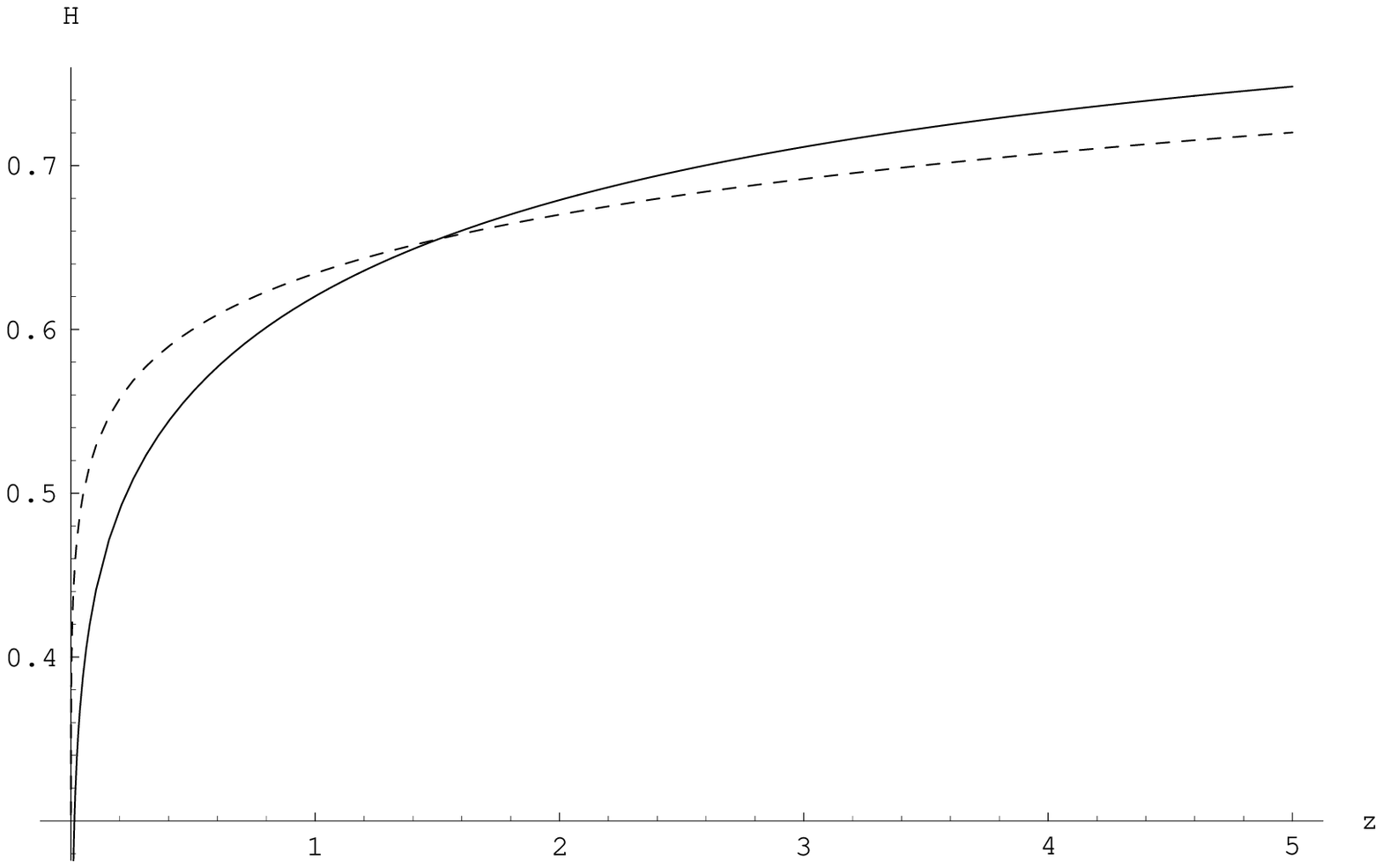,width=5in}}} 
\end{picture}  
\caption{A comparison of the transformed weight with a form
obtained by scaling from high energy (dashed curve), as described in
the text. Here the power is $\delta = 0.92$. } 
\end{figure}

The closeness of these exponents to unity suggests that we try to
accommodate a behavior $\rho (m) \sim 1/m$. Surely a weight function of
this inverse power is a more pleasing result than one with a
non-integral power. However, a pure power-law with $\delta=1$ is not
integrable at the low energy end. We can form a normalizable weight if
we include a low-energy cutoff.
\begin{equation}
\rho_3(m) = {N \over m} \theta (m - m_{MIN}) \theta ( m^* - m)
\end{equation}
with $N = 1/ ln(m^*/m_{MIN})$. At this stage the origin of the
low-energy cutoff is unexplained; however, fortunately $m_{MIN}$
enters only logarithmically. Fig 17 displays these forms for values 
$m_{MIN} = m_e$ and $m_e/100$. We see that these forms in
fact do very well at describing the transformed weight within the
intrinsic uncertainty. 
\begin{figure}  
\unitlength0.05in  
\begin{picture}(100,75)  
\put(0,0){\makebox(100,75)  
{\epsfig{figure=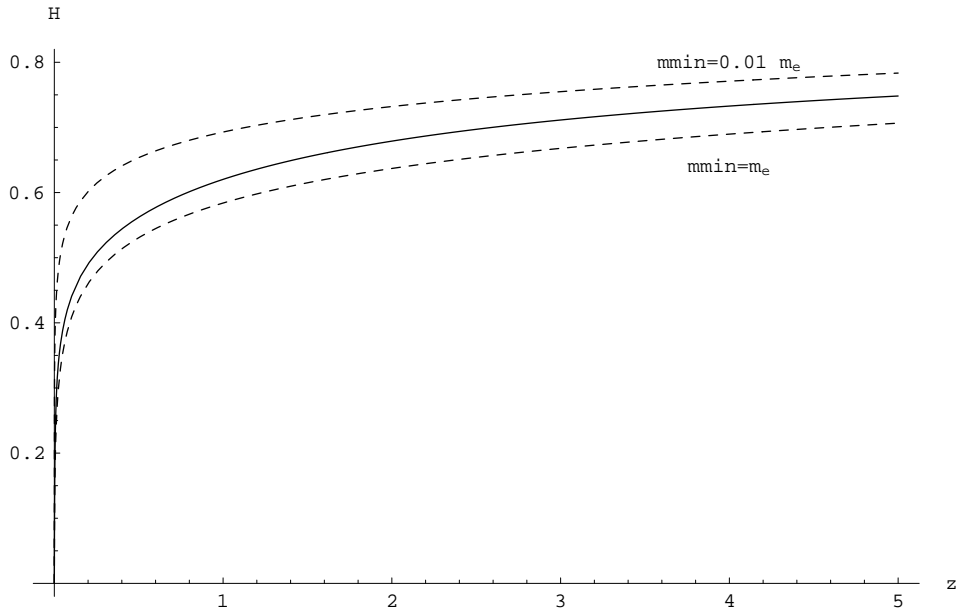,width=5in}}} 
\end{picture}  
\caption{ The transformed weight compared with scale-invariant
forms with cut-offs $m_e$ and $m_e/100$. } 
\end{figure}  

For this form of the weight, the median value of the distribution is
determined by the endpoints in the simple form
\begin{equation}
\hat{m} = \sqrt{m^*~m_{MIN}} 
\end{equation}
For $m_{MIN} = m_e$, this equals $\hat{m} = 0.34$ GeV, a reasonable
value. 

The weights with $\rho (m) \sim 1/m$ can be described as ``scale
invariant'' in the following senses. In the first place, there is no
scale (other than the endpoints) in the shape of $\rho (m)$, and the
normalization constant is dimensionless and 
independent of the overall scale. In
addition, under any linear rescaling of the masses such as
\begin{equation}
m_2(\mu_2) = \left( {\alpha_s(\mu_2) \over \alpha_s(mu_1)}
\right)^{d_m}~   m_1(\mu_1)
\end{equation}
the transformation rule of Eq. 5 tells us that this weight will remain
unchanged (again, aside from the endpoints), since
\begin{eqnarray}
\rho_\mu (m) & = & \rho_{\mu_1}\left( m \left({\alpha_s(\mu_) \over
\alpha_s(mu_1)} \right)^{-d_m}\right) \left({\alpha_s(\mu_) 
\over \alpha_s(mu_1)}\right)^{-d_m}  \\
& = &  {1 \over  m \left({\alpha_s(\mu_) \over
\alpha_s(mu_1)} \right)^{-d_m}} \left({\alpha_s(\mu_)
\over \alpha_s(mu_1)}\right)^{-d_m} \\
& = & {1 \over m} .
\end{eqnarray}
It is tempting to speculate that this scale invariant form could
approximate an IR fixed point for some class of weights initially
defined at high energy.

The best fit functional form for the weight is shown in Fig 18 and 19.
It is the scale-invariant form with $m_{MIN}=m_e / 4$, and tracks the
central ``experimental'' curve quite closely. However, the specifics
of this form need not be taken too seriously given the uncertainties
inherent in this problem. Many of the trial functions shown above are
compatible within the uncertainty. However, all of these forms are
strikingly close to $\delta = 1$, suggesting that this is the key to
the structure of the mass spectrum.
\begin{figure}  
\unitlength0.05in  
\begin{picture}(100,75)  
\put(0,0){\makebox(100,75)  
{\epsfig{figure=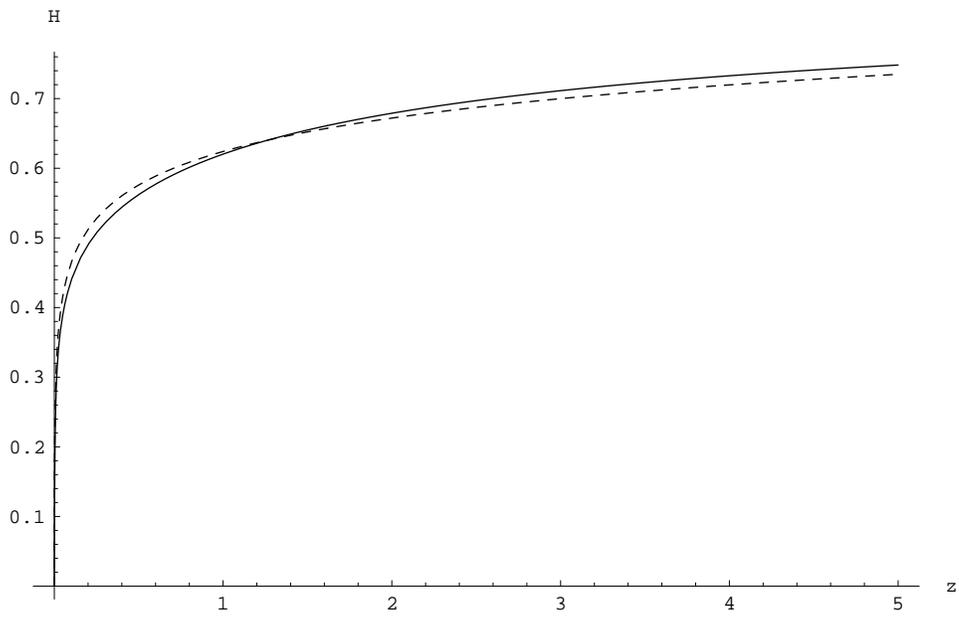,width=5in}}} 
\end{picture}  
\caption{The best description of the transformed weight,
obtained with the scale-invariant form using $m_{MIN} = 0.4 m_e$. } 
\end{figure}  
\begin{figure}  
\unitlength0.05in  
\begin{picture}(100,75)  
\put(0,0){\makebox(100,75)  
{\epsfig{figure=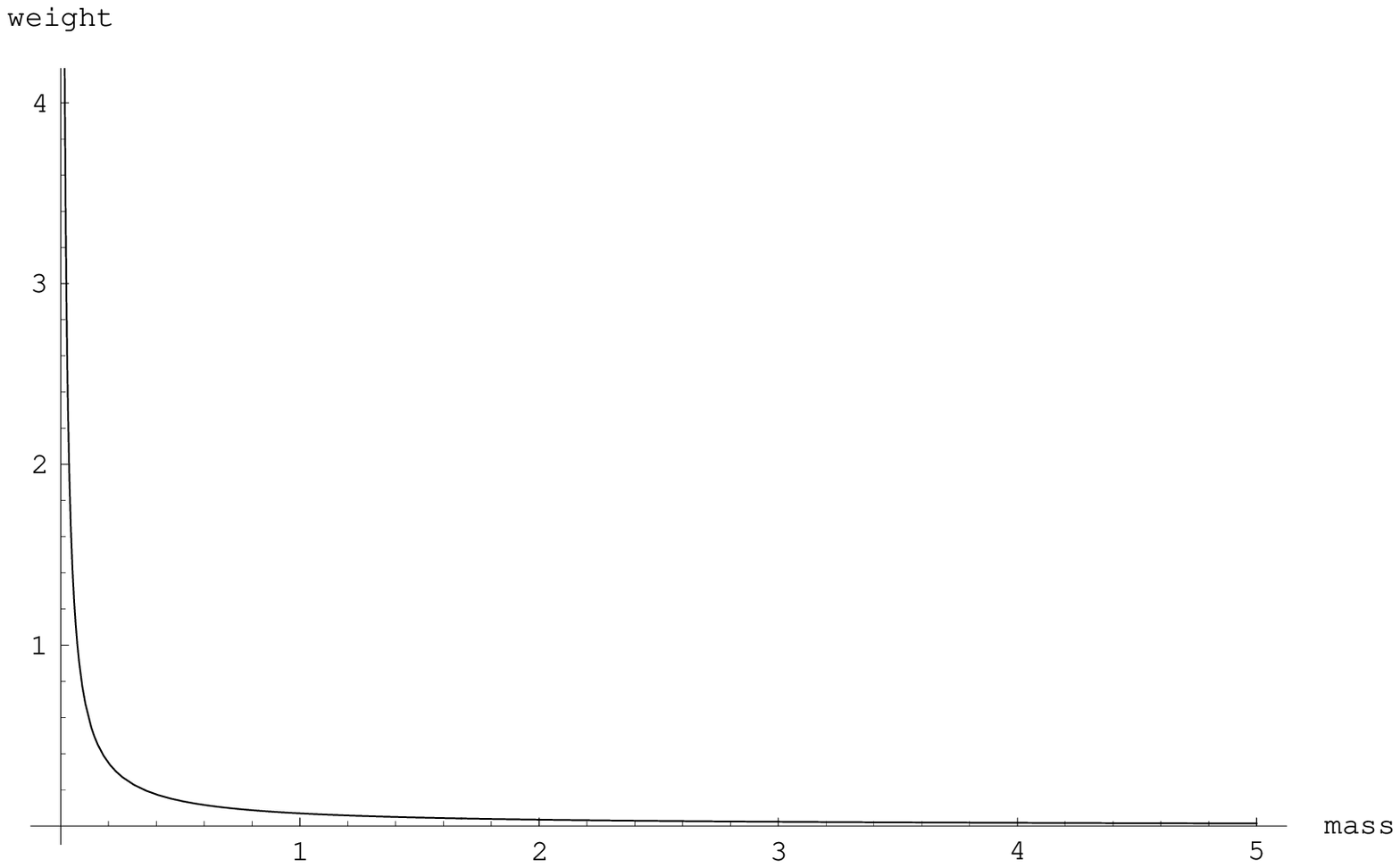,width=5in}}} 
\end{picture}  
\caption{The scale invariant weight describing the best
match to the quark mass spectrum, corresponding to Fig. 18. } 
\end{figure}

\section{Further comments and summary}

  The previous sections have contained a first consideration of the
``experiment'' and phenomenology of the weight function for quark
masses. The hope is that this weight is the visible remnant of the
fundamental theory in situations where the specific values of the
masses are themselves not unique. 

There are some considerations which could be important in trying to
predict the weight function. In certain cases, it may prove that a
slightly modified ``experimental'' description is most relevant for a
given theory. The procedure used here is specific to 
the weight in a domain
with 3 generations of fermions in an $SU(3)\times SU(2) \times U(1)$
theory with the observeredcalues of the gauge couplings. 
These other features may potentially also
be variable, and the weight function could be different for other
situations. In the case that the Higgs vacuum expectation value is
also variable, it is simple to convert $\rho (m)$ into a
distribution of the Yukawa couplings. While there are a few subtle
features of transforming the Yukawa distribution to other scales,
these are small in comparison with the intrinsic uncertainty in the
distribution. 

The procedure used above implicitly assumes that it is the quark
masses themselves that are independently distributed with respect to
some weight. In specific models, this may not be the case. For example
in models with an intrinsic hierarchy between different Yukawa
couplings generated through radiative corrections, 
it may turn out that the smallness of some masses is the result of
high powers of a gauge coupling rather than a consequence of the
weight itself. In this case, a different procedure to extract the
weight for the appropriate random variables would need to be employed.

Of more serious concern could be a bias introduced by anthropic
considerations. In multiple domain theories, it is a natural
requirement that out of all the possible domains we could only find
ourselves in a domain that has the ingredients relevant for life.
Without too much anthro-centric reasoning, it seems plausible that
this requirement implies the need for complex chemicals (i.e. more
elements than simply hydrogen). In [4], this was argued to require
that at
least some of the quark masses must be small compared to the QCD scale.
This forms a bias for low masses, and would shift the shape of the
weight function. The ``experimental'' weight function must then be
understood to be subject to this constraint.

The issues described above are best treated in the context of a theory
in which we try to predict the weight. Depending on the dynamics of
this hypothetical theory, we would best know which variables are
quasi-random and over what range. Within the parameter space of the
theory, we can impose constraints on other parameters and explore the
distribution of masses subject to those constraint. It is even possible
that we could estimate the effects of the anthropic
bias, producing a weight function subject to the constraint that
complex chemicals are able to be formed. It would be interesting to
explore these issues even in the context of a toy model.

In theories where the parameters are variable in different domains of
the universe, many of the standard questions that we address in
particle physics appear in a different light. 
In some cases, such as the attempt 
to understand the scale of electroweak symmetry breaking, the form of
analysis has a quite different character[4]. For quasi-random quark
masses, we might fear that there is in this case no longer anything
that we can do phenomenologically with the masses. This paper has 
been an attempt to extract the remnant of the underlying theory which
survives even in the case where the specific values of the masses are
not unique. The result is intriguingly close to a scale-invariant
weight (see Figs. 18 and 19), and 
this encodes the observed bias for small quark masses.
This form of theory is relatively new, and it is not clear how much it
will be developed in the future. To the extent that these theories are
studied further, the weight for fermion masses will be a fundamental
input that has the potential to test the theories.   
 
\section*{Acknowledgments} 

I would like to thank Gene Golowich, Dave Seckel and especially Steve
Barr for conversations and communications which have helped shape my
thoughts on these unusual topics.  

\section*{References}

\begin{enumerate}
\item A. Linde, {\it Phys. Lett.} {\bf B129}, 177 (1983);
{\it ibid.} {\bf B175}, 395 (1986), {\it ibid.} {\bf B202},
194 (1988), Phys. Scri.,{\bf T15}, 169 (1987).
\item See eg. M. Green, J. Schwartz and E. Witten, {\it Superstrings}
(Cambridge University Press, Cambridge, 1987), and
G. Ross in {\it CP Violation and the Limits of the Standard
Model - TASI-94}, ed by J. F. Donoghue (World Scientific, Singapore, 1995). 
\item S. Weinberg, {\it Phys. Rev. Lett.} {\bf 59}, 2607 (1987); \\
S. Weinberg, astro-ph/9610044, {\it Critical Dialogues in Cosmology} 
ed. by J. Bachal and N. Turok (World Scientific, Singapore, 1997);  \\
H. Martel, P. Shapiro, and S. Weinberg, astro-ph/9701099.
\item V. Agrawal, S. M. Barr, J. F. Donoghue and D. Seckel, hep-ph/9707380, 
to be published.
\item B. Carter, in I.A.U. Symposium, Vol 63, ed by M. Longair (Reidel,
Dordrecht, 1974); \\
J. Barrow and F. Tipler, {\it The Anthropic Cosmological Principle}
(Clarendon Press, Oxford, 1986); \\
B. J. Carr and M.J. Rees, Nature {\bf 278}, 605 (1979); \\
A. Vilenkin, Phys. Rev. Lett. {\bf 74}, 846 (1995).
\item B. Pendelton and G. Ross, Phys. Lett. {\bf B98}, 291 (1981); \\
C. T. Hill, Phys. Rev. {\bf D24}, 691 (1981);  \\
C. T. Hill, C. N. Leung and S. Rao, Nucl. Phys. {\bf B262}, 517 (1985).
\item G. Ross and M. Lanzagorta, Phys. Lett {\bf D349}, 319 (1995).
\item Review of Particle Properties, R.M. Barnett et.al., Phys. Rev.
{\bf D54}, 1 (1996).
\item J. Gasser and H. Leutwyler, Physics Reports {\bf 87C}, 77 (1982);\\
J. F. Donoghue, E. Golowich and B. R. Holstein, {\it Dynamics of the
Standard Model}, (Cambridge University Press, Cambridge, 1992). 
\item see the review and references in M. Neubert, hep-ph/9610266, and
hep-ph/9404296, published in {\it The Building Blocks of Creation
- From Microfermis to Megaparsecs -TASI-94} , ed by S. Raby and T.
Walker (World
Scientific, Singapore, 1994).
\end{enumerate}

\end{document}